\documentclass[a4paper,11pt]{article}
\usepackage[a4paper, total={17cm, 25cm}]{geometry}
\usepackage{graphicx}
\usepackage{float}
\usepackage{fancyhdr}
\fancypagestyle{plain}{
\fancyhf{} 
\fancyhead[RH]{\thepage}
\fancyfoot[CF]{28th Texas Symposium on Relativistic Astrophysics \\ Geneva, Switzerland -- December 13-18, 2015}

}
\begin{document}

\title{\bf Coeval Observations of a Complete Sample of Blazars with Effelsberg, IRAM 30m, and Planck}

\author{
J\"org P. Rachen%
\footnote{j.rachen@astro.ru.nl}%
,\\ Department of Astrophysics\,/\,IMAPP, Radboud University Nijmegen, The Netherlands;\\ previously at: Max-Planck-Institute for Astrophysics, Garching, Germany.
\\[3pt]%
Lars Fuhrmann%
\footnote{present address: ZESS - Center for Sensorsystems, University of Siegen, Germany.}%
, Thomas Krichbaum, Emmanouil Angelakis, Ioannis Nestoras, Anton Zensus,\\ Max-Planck-Institute for Radioastronomy (MPIfR), Bonn, Germany.
\\[3pt]%
Albrecht Sievers, Hans Ungerechts,\\ Instituto de Radio Astronom\'ia Milim\'etrica (IRAM), Granada, Spain.
\\[3pt]%
Elina Keih\"anen,\\ Department of Physics, University of Helsinki, Finland.
\\[3pt]%
Martin Reinecke,\\ Max-Planck-Institute for Astrophysics, Garching, Germany. 
}

\date{}

\maketitle

\vspace{-17pt}

\begin{abstract}
\noindent  We present the outline and first results of a project using the synergies of the long term blazar radio-millimetre monitoring program
{\sc\large f-gamma}, the continued scanning of the millimetre-submillimetre sky by the \textit{Planck} satellite, together with several dedicated
observing programs at the Effelsberg 100m telescope, to obtain a data sample unprecedented in both time resolution and frequency span.
\end{abstract}


\section{Introduction}

Highly time resolved multifrequency observations of blazars are the key to our understanding of the extreme physics of supermassive black holes and their jets, 
called active galactic nuclei (AGN). A particular role play here the non-thermal fringes of the electromagnetic spectrum, the radio and $\gamma$-ray regime, where 
strongly variable emission from relativistic jets directed towards the observer dominates the AGN spectrum. Since the launch of the {\it Fermi}\/-GST in June 2008,
the correlation of the emission of these regimes has been the subject of intensive research, most prominently with the {\it Fermi}\/-GST related \mbox{\sc\large f-gamma} program 
\cite{Fuhrmann2007, Fuhrmann2014}. Less investigated is for such highly variable sources the submillimetre regime, marking the transition to the thermal emission of AGN. Here, the 
{\it Planck} satellite, built to investigate the structure of the cosmic microwave background \cite{Tauber2010,Planck-I-2011}, provides a valuable data set, 
which partly overlaps the wavebands covered by the {\sc\large f-gamma} program, but extends the range to 857\,GHz ($350\,\mu$m). This can give insight on possible submillimetre 
variability and its relation to classical radio variability, and set the path for future observations and monitoring involving instruments like ALMA, covering the same frequency range 
as {\it Planck} with significantly higher sensitivity and resolution. 

\section{Targets and observations}

\begin{table}[b!p]

\newcommand{\fm}[1][]{\footnotemark[#1]}
\renewcommand{\thefootnote}{\fnsymbol{footnote}}
\setlength{\tabcolsep}{4.9pt}

\begin{footnotesize}

\caption[]{\small The Planck-Effelsberg complete sample. The selected common names will be used to refer to sources in analysis and discussion. Fluxes at 8.4 GHz and 143 GHz, 
taken from {\sc crates } and \textit{Planck} {\sc ercsc}, respectively, are given in Jy, $z$ denotes the source redshift.} 

\vspace{7pt}

\begin{tabular}{llrll||llrrl}
{\sc crates} name 		& common name &  	$F_{8.4}$ 	& $F_{143}$ 	& $z$ 	& {\sc crates} name 				& common name &  	$F_{8.4}$ 	& $F_{143}$ 	& $z$ 	\\\hline
J010838${+}$013516		& 4C\,${+}$01.02	& 2.26	& 1.63		& 2.10	& J122906${+}$020245\fm[1]		& 3C\,273			& 41.73	& 9.58		& 0.54	\\
J021731${+}$734935		& S5\,0212${+}$73	& 2.29	& 1.05		& 2.37	& J135704${+}$191919			& 4C\,${+}$19.44	& 1.13	& 0.87		& 0.72	\\
J023752${+}$284814\fm[1]	& 4C\,${+}$28.07	& 2.76	& 1.06		& 1.21	& J155035${+}$052702			& 4C\,${+}$05.64	& 1.61	& 1.06		& 1.44	\\
J031947${+}$413042\fm[1]	& 3C\,84			& 28.01	& 7.29		& 0.02	& J163515${+}$380813\fm[1]		& 4C\,${+}$38.41	& 2.40	& 2.66		& 1.81	\\
J032152${+}$122123		& PKS\,0319${+}$12	& 1.11	& 0.76\fm[3]	& 2.66	& J163813${+}$572029			& TXS\,1637${+}$574	& 1.34	& 0.70		& 0.75	\\
J033930${-}$014638		& PKS\,0336${-}$01	& 2.69	& 0.83		& 0.85	& J164207${+}$685647\fm[2]		& 4C\,${+}$69.21	& 1.21	& 1.25		& 0.75	\\
J042315${-}$012034\fm[1]	& PKS\,0420${-}$01	& 2.41	& 5.27		& 0.91	& J164258${+}$394842\fm[1]		& 3C\,345			& 6.30	& 4.05		& 0.59	\\
J043310${+}$052115		& 3C\,120			& 2.11	& 1.48		& 0.03	& J180045${+}$782804\fm[1]\fm[2]	& S5\,1803${+}$78	& 2.87	& 1.59		& 0.68	\\
J053238${+}$073238		& TXS\,0529${+}$075	& 2.76	& 0.77		& 1.25	& J180651${+}$694931\fm[2]		& 3C\,371			& 1.60	& 0.90		& 0.05	\\
J060800${-}$083454		& PKS\,0605${-}$08	& 1.97	& 1.24		& 0.87	& J182406${+}$565059\fm[2]		& 4C\,${+}$56.27	& 1.19	& 0.99		& 0.66	\\
J073917${+}$013656		& PKS\,0736${+}$01	& 1.71	& 1.27		& 0.19	& J192747${+}$735755\fm[2]		& 4C\,${+}$73.18	& 3.70	& 2.42		& 0.30	\\
J075051${+}$123113		& PKS\,0748${+}$126	& 1.97	& 2.04		& 0.89	& J203154${+}$121929			& PKS\,2029${+}$121	& 1.10	& 0.86	  	& 1.22	\\
J080815${-}$075109		& PKS\,0805${-}$07	& 1.76	& 1.72		& 1.84	& J220315${+}$314538			& 4C\,${+}$31.63	& 2.75	& 2.04		& 0.30	\\
J081816${+}$422248\fm[1]	& TXS\,0814${+}$425	& 1.04	& 0.95		& 0.53	& J222546${-}$045700			& 3C\,446			& 3.25	& 2.54		& 1.40	\\
J105830${+}$013340		& 4C\,${+}$01.28	& 3.36	& 2.92		& 0.89	& J222940${-}$083254			& PKS\,2227${-}$08	& 1.23	& 1.56		& 1.60	\\
J122221${+}$041316		& 4C\,${+}$04.42	& 1.02	& 0.99		& 0.97	& J225358${+}$160853\fm[1]		& 3C\,454.3		& 10.38	& 27.94		& 0.86	\\\hline
\multicolumn{5}{l}{{\Large\strut}\fm[1] {\sc\small f-gamma} source} & \multicolumn{5}{l}{\fm[2] Planck deep field}\\
\multicolumn{10}{l}{\fm[3] Flux taken from \textit{Planck} {\sc\large ercsc} 217\,GHz catalog, as source was not contained in the 143\,GHz catalog.}\\
\end{tabular}
\end{footnotesize}

\end{table}

\subsection{F-GAMMA observations}

The \textit{Fermi}\/-GST related monitoring program of gamma-ray blazars (\mbox{\sc\large f-gamma}) closely coordinated Effelsberg 100m observations with the
more general flux monitoring at the IRAM 30m telescope between 2008 and 2014. Effelsberg measurements were conducted with the secondary focus heterodyne
receivers at $2.64$, $4.85$, $8.35$, $10.45$, $14.60$, $23.05$, $32.00$ and $43.00$\,GHz, performed quasi-simultaneously with cross-scans, slewing over the
source position in azimuth and elevation direction with an adaptive number of sub-scans \cite{Fuhrmann2007, Fuhrmann2014, Angelakis2015}. The IRAM 30m observations
were carried out with calibrated cross-scans using the EMIR bands at 86.2 and 142.3 GHz in horizontal and vertical polarization.  The opacity corrected 
intensities were converted into the standard temperature scale and finally corrected for pointing offsets and systematic gain-elevation effects. The conversion to standard 
flux densities was done using frequently observed primary (Mars, Uranus) and secondary (W3(OH), K3-50A, NGC\,7027) calibrators.
The \mbox{\sc\large f-gamma} target sample is based on a collection of about 90 \textit{Fermi}-bright blazars, of which 60 have been regularly monitored with a cadence of 
about 1 month. Subsamples of these have been scientifically analysed in Refs.~\cite{Fuhrmann2014, Angelakis2015}, where also a full description of the sample can be found.

\subsection{Complete sample of flat spectrum blazars and other targets}

To allow solid statistical analysis of combined Planck-Effelsberg data, we have defined a complete sample of flat spectrum blazars derived from the 
{\sc\large crates} catalog \cite{CRATES-2007}, applying the spectral index cuts $\alpha_{\rm <4.8GHz} > -0.3$ and $\alpha_{\rm <8.4GHz}\,{<}\,0.1$ (convention: 
$F_\nu \propto \nu^{-\alpha}$), a flux limit $F_{\rm 8.4GHz}\,{>}\,1.0\,$Jy, and declination $\delta>-10^\circ$. These criteria focus on sources with flat cm-spectra, but exclude 
GPS objects. To ensure observability by \textit{Planck}, we further required $F_{\rm \ge 143GHz}>700\,$mJy, for which we used the \textit{Planck} {\sc\large ercsc} at $143$ and $217$\,GHz 
\cite{Planck-VII-2011}. The total sample of 32 sources is listed in Table 1. It has an overlap of 9 sources with the core \mbox{\sc\large f-gamma} sample, and 26 sources with the 
Mets\"ahovi sample of northern radio sources otherwise used for blazar research with \textit{Planck} \cite{Planck-XV-2011}. Five sources are located in the so-called \textit{Planck deep field} 
around the ecliptic north-pole, which are much more frequently hit by \textit{Planck} than other sources of the sky \cite{Tauber2010,Planck-I-2011}. Because of the importance 
of sources in this region for analysis of \textit{Planck} measured flux variations on short time scales (see below), we have observed two more deep field blazars, S4\,1749${+}$70 
and S4\,1849${+}$67, which will be individually analysed but not included in statistical analysis. All sources in this target list were observed with the Effelsberg 100m telescope 
between April 2011 and February 2012 with a cadence of about one month and a two week offset from \mbox{\sc\large f-gamma} observations. Observations of individual 
sources have been restricted to times maximally 2 months from a \textit{Planck} scan, thus following \textit{Planck} during its 4$^{\rm th}$ and 5$^{\rm th}$ sky survey. Most observations 
were using five Effelsberg receivers at $4.85$, $8.35$, $14.60$, $23.05$, and $32.00$ GHz, but some were performed using all the receivers used by the \mbox{\sc\large f-gamma} program.

\subsection{Time-resolved Planck fluxes}

In most \textit{Planck} related analysis of CMB foregrounds, point sources are extracted from time-restricted sky maps \cite[and subsequent work]{Planck-XV-2011, Giommi2012, Chen2013}. 
Such extraction methods are well established and validated, but restricted to pre-defined time cuts of the available maps, and potentially subject to artifacts in the 
sources spectra at the edges of survey cuts \cite{Planck-XIV-2011}. To overcome this restriction, \textit{Planck} fluxes used here are directly extracted from the \textit{Planck} time ordered information (TOI),
utilising tools developed for blind variability searches in \textit{Planck} data \cite{Rachen2015} based on the beam-deconvolution code ArtDeco \cite{Keihanen2012}. 
These tools detect flux \textit{variations} in a given sky position and associate them with bright variable sources within the beam. For every pointing ID (PID) and each detector, 
a source flux is estimated by de-convolving with the detector beam to the position of the source. While sweeping over the source, \textit{Planck} produces a ``survey scan'' 
profile with roughly one hour time resolution. Such ``single PID flux deviation estimates'' suffer stongly from instrument noise, systematics due to our incomplete knowledge of scanning 
beam profiles, precise pointing, and gain variations, plus technical limitations of the method applied, and thus show a large scattering between individual PIDs. To obtain reasonable 
flux estimates, they are therfore combined to single-survey fluxes, which represent the average deviations from the mean flux over several hours to one day, with errors conservatively 
taken as the standard deviation of individual PID fluxes over the survey scan. Total fluxes are determined by adding the measured average deviation to the source flux listed in the 
Second Planck Catalogue of Compact Sources \cite{Planck-XXVI-2015}, which contains fluxes of point sources averaged over the entire mission. Figure 1 shows single survey fluxes for 
the bight blazar 3C84 together with \mbox{\sc\large f-gamma} monitoring data. More options are given for blazars in the \textit{Planck deep field}, where single PID results can be combined 
according to signal-to-noise requirements to long time lines at selected frequencies.

\begin{figure}

\includegraphics[width=1.023\textwidth,viewport=23 63 787 542,clip=true]{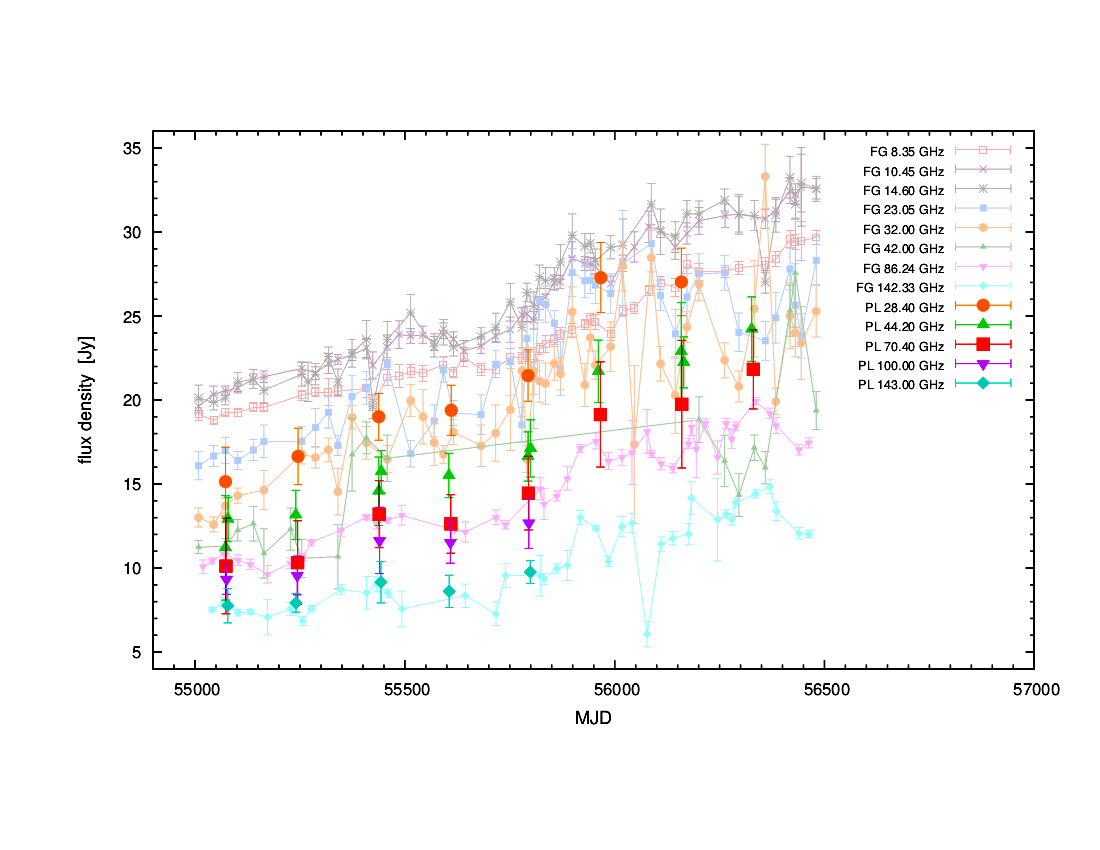}

\caption[]{\footnotesize
{\sc\normalsize f-gamma} light curves from 8.35 to 142.3 GHz for the bright blazar 3C84 between Aug 2009 and Dec 2013, with fluxes of the Planck LFI 30, 44, and 70 GHz radiometers, 
HFI 100 GHz PSBs and HFI 143 GHz SWBs, noted at their central bandpass frequencies \cite[and references therein]{Tauber2010,Planck-I-2011}.  \textit{Planck} fluxes are derived from TOI mapping 
as weighted averages over full scans and all detectors at one frequency combined, errors express the rms scattering of the single PID fluxes during each scan (see text and Ref.~\cite{Rachen2015}).  
}

\end{figure}

\section{Project goals}

\vspace{-2.3pt}

\subsection{Radio to submillimetre monitoring of blazars on long and short time scales}

\textit{Planck} time resolved fluxes combined with cm-mm data can add to our knowledge on blazar variability in several ways: First, by determining quasi-simultaneous spectra extending 
over $2.5$ orders of magnitude in frequency, where \textit{Planck} refines and extends the frequency coverage, as 
demonstrated in Refs. \cite{Planck-XV-2011, Chen2013, Planck-XIV-2011}. Second, for sources located in the \textit{Planck} deep field it allows, albeit in irregular patters, monitoring 
at single frequencies on sub-day-scale resolution. A tentative detection of a flare with one week time scale in the blazar S5\,1803${+}$78 at 30\,GHz (starting April 2, 2012), which was 
entirely missed by the {\sc\large f-gamma} monitoring, might be a first result in this direction, but still requires checks for instrumental systematics. The potential of \textit{Planck} for such 
investigation is given in particular for frequencies with large beamwidths (30 and 44 GHz) or covering several detector rows in the \textit{Planck} focal plane (70, 143, 217 GHz). A particular case is the Planck 44 GHz channel, which has detectors at both ends of the focal plane and thus allows conclusions on one-week scale variability for essentially all sources in the sky (see Refs.~\cite{Planck-XIV-2011,Tauber2010-II} for details). 

\vspace{-2.3pt}

\subsection{External validation of Planck variability analysis and project time line}

Another important role of this project is the external validation of the \textit{Planck} blind variability detection method \cite{Rachen2015}, as it allows to check the reliability of detected flux variations
on PID time scales and the discrimination from instrumental effects by comparing TOI-derived fluxes with quasi simultaneous measuremets with external telescopes. First comparisons made 
for the nine {\sc\large f-gamma} sources in our complete sample have shown an excellent agreement between HFI 143 GHz SWB fluxes and IRAM\,30m 143\,GHz, while the comparison of LFI 30\,GHz 
with Effelsberg 32\,GHz shows systematically higher \textit{Planck} fluxes, which may be explained by the offset in center frequencies, the very different beam sizes of \textit{Planck} and Effelsberg, or 
methodological systematics. Further investigations are underway, and we expect that both the blind search for variability and the time resolved flux extraction will be finished well before the final 
release of \textit{Planck} results.

\vspace{-2.3pt}

\subsection*{Acknowledgements}

\begin{small}

This project makes use of observations with the 100m telescope of the MPIfR at Effelsberg and with the IRAM 30m telescope on Pico Veleta. 
We acknowledge support by the {\sc\normalsize f-gamma} team and the Planck Collaboration, in particular by Fran\c{c}ois Bouchet, Carlo Burigani, Torsten En{\ss}lin, Ken Ganga, 
Reijo Keskatilo, Marcos L\'opez-Caniego, \mbox{Nazzareno} Mandolesi, Marcella Massardi, Sylvain Mottet, Bruce Partridge, and Andrea Zacchei. 


\end{small}


\vspace{-2.3pt}

\begin{thebibliography}{99}
\setlength{\itemsep}{0pt}

\bibitem{Fuhrmann2007} Fuhrmann, L., Zensus, J.A., Krichbaum, T.P., Angelakis,~E., Readhead, A.C.S.,
AIPC 921, 249 (2007)

\bibitem{Fuhrmann2014} Fuhrmann, L., Larsson, S., Chiang, J., et al., \mbox{MNRAS 441}, 1899 (2014)

\bibitem{Tauber2010} Tauber, J.A., Mandolesi, N., Puget J.-L., et al., A\&A 520, A1 (2010)

\bibitem{Planck-I-2011} Planck Collaboration I, A\&A 536, A1 (2011)

\bibitem{Angelakis2015} Angelakis, E., Fuhrmann, L., Marchili, N., et al., \mbox{A\&A 575}, A55 (2015)

\bibitem{CRATES-2007} Healey, S.E., Romani, R.W., Taylor, G.B., et al., \mbox{ApJS 171}, 61 (2007) 

\bibitem{Planck-VII-2011} Planck Collaboration VII, A\&A 536, A7 (2011)

\bibitem{Planck-XV-2011} Planck Collaboration XV, A\&A 536, A15 (2011)

\bibitem{Giommi2012} Giommi, P., Polenta, G., L\"ahteenm\"aki, A., et al., A\&A 541, A160 (2012)

\bibitem{Chen2013} Chen, X., Rachen, J.P., L\'opez-Caniego, M., et al., \mbox{A\&A 553}, A107 (2013)

\bibitem{Planck-XIV-2011} Planck Collaboration XIV, A\&A 536, A14 (2011)

\bibitem{Rachen2015} Rachen, J.P., Keih\"anen, E., Reinecke, M, arXiv1512.03737 (2015)

\bibitem{Keihanen2012} Keih\"anen, E., Reinecke, M., A\&A 548, A110 (2012)

\bibitem{Planck-XXVI-2015} Planck Collaboration XXVI, arXiv1507.02058 (2015)

\bibitem{Tauber2010-II} Tauber, J.A., Noorgard-Nielsen, H.U., et al., A\&A 520, A2 (2010)

\end{thebibliography}
\end{document}